\documentclass[showpacs,prb,letterpaper,twocolumn,floatfix,nobalancelastpage]{revtex4}

\usepackage{amsmath}
\usepackage{bm}
\usepackage{graphicx}
\usepackage{color}
\usepackage{float}
\begin{document}

%
\def\bra#1{\mathinner{\langle{#1}|}}
\def\ket#1{\mathinner{|{#1}\rangle}}
\def\braket#1{\mathinner{\langle{#1}\rangle}}
\def\Bra#1{\left<#1\right|}
\def\Ket#1{\left|#1\right>}
{\catcode`\|=\active
  \gdef\Braket#1{\left<\mathcode`\|"8000\let|\BraVert {#1}\right>}}
\def\BraVert{\egroup\,\mid@vertical\,\bgroup}
{\catcode`\|=\active
  \gdef\set#1{\mathinner{\lbrace\,{\mathcode`\|"8000\let|\midvert #1}\,\rbrace}}
  \gdef\Set#1{\left\{\:{\mathcode`\|"8000\let|\SetVert #1}\:\right\}}}
\def\midvert{\egroup\mid\bgroup}
\def\SetVert{\egroup\;\mid@vertical\;\bgroup}

\def\bra#1{\mathinner{\langle{#1}|}}
\def\ket#1{\mathinner{|{#1}\rangle}}
\def\braket#1{\mathinner{\langle{#1}\rangle}}
\newcommand{\G}[1][{}]{\ensuremath{{{\mathbf G}}_\mathrm{#1}}}

\title{Influence of electron-acoustic phonon scattering
on off-resonant cavity feeding within a
strongly coupled quantum-dot  cavity system}
\author{S. Hughes}
\email{shughes@physics.queensu.ca}
\author{P. Yao}
\address{Department of Physics, Queen's University\\
Kingston, ON  K7L 3N6 Canada}
\author{F. Milde and A. Knorr}
\address{
Institut f\"ur Theoretische Physik, Nichtlineare Optik und Quantenelektronik, Technische Universit\"at Berlin, Hardenbergstra\ss e 36,
PN 7-1 10623 Berlin, Germany}
\author{D. Dalacu, K. Mnaymneh, V. Sazonova, P. J. Poole, G. C. Aers, J. Lapointe, R. Cheriton,
and R. L. Williams}
\address{Institute for Microstructural Sciences, National Research Council, Ottawa, Canada K1A 0R6}

\begin{abstract}
We present  a
medium-dependent
quantum optics approach to describe the influence
of electron-acoustic phonon coupling on the emission spectra
of a strongly coupled quantum-dot  cavity system.
Using a canonical Hamiltonian for light quantization and
a  photon Green function formalism,
phonons are included to all orders
through the dot polarizability function obtained
within  the independent Boson model.
We
derive simple user-friendly analytical expressions for the linear quantum light spectrum,
including the influence from both  exciton and cavity-emission decay channels.
In the regime of semiconductor cavity-QED, we  study cavity emission for various exciton-cavity detunings and
demonstrate rich spectral asymmetries as well
as
cavity-mode suppression and enhancement effects.
Our
technique
is nonperturbative, and non-Markovian,  and
can be applied to study photon emission from a wide range of semiconductor quantum dot  structures,
including waveguides and coupled cavity arrays.
 We compare our theory
directly to recent and apparently puzzling experimental data for a single site-controlled quantum dot
in a photonic crystal cavity and
show  good agreement as a function of cavity-dot detuning and as a function of temperature.

\end{abstract}

\pacs{42.50.Ct,  78.67.Hc, 42.50.Pq}

\maketitle

\section{Introduction}

The influence of electron-acoustic phonon scattering is a well known effect
in semiconductor quantum dot (QDs).
Due to the interplay of phonon emission and absorption,
the longitudinal acoustic (LA) phonon bath  manifests in spectral lineshapes that are highly asymmetric at low temperature,
and sit on the background of the symmetric zero phonon line (ZPL).
The characteristic spectral lineshape of LA phonon
scattering on electron-hole pairs (``excitons'')
in quasi-homogeneous semiconductor structures
is well described through the independent Boson model (IBM)
\cite{MahanBook:1981,KrummheuerPRB:2001,VagovPRB:2002,KnorrPRL:2003},
 and IBM simulations have shown
 good agreement with experiments.
 In addition to the IBM lineshape,
 the broadening of the ZPL
 is usually described phenomenologically,  though this can be reliably fit to experiments \cite{BorriPRL:2001}; however, there is still some controversy as to the origin of the
 ZPL broadening which can include contributions from
radiative broadening, spectral diffusion, anharmonicity effects \cite{ZimmermannPRL:2004,MachnikowskiPRL:2006}, phonon scattering
from interfaces \cite{OrtnerPRB:2004,RudinPRB:2006}, and a modified phonon spectrum \cite{KnorrPRL:2007}.
While it is well known that the phonons cause the exciton lineshapes to be highly
non-Lorentzian (stemming from non-Markovian decay),
most of the present QD cavity-QED theories
only add in a Lorentzian broadening mechanism for the QD excitons, e.g., at the
level of a Markovian master equation.

Recently, there have been several studies of the role
of electron-phonon coupling in semicondutor QD cavity
systems. Wilson-Rae and Imamogl{\u u}  \cite{WilsonRayPRB:2004} treated the phonon interaction
with QDs using polaron Green functions and derived an {\em analytic}
absorption lineshape when the dot and cavity are on resonance; in the polaron
representation, new phonon-induced interaction terms are introduced exactly,
while a second-order Born approximation was applied to
include ``residual''
exciton-photon-phonon coupling effects.
Polaron approaches have also been recently employed
to successfully describe phonon-induced decay of optical pulse-excited QDs in the absence of any cavity coupling \cite{NazirPRL:2010}.
Milde {\em et al.} \cite{MildePRB:2007} numerically solved the IBM and coupled
the QD susceptibility to a photonic crystal cavity system through a semiclassical Green function
approach, demonstrating
asymmetries in the on-resonance Rabi doublet
and the effect of increasing temperature for both the cavity
emitted spectra and the side-coupled waveguide transmission.
Xue {\em et al.}~\cite{JiaoJPC2008} have  applied perturbation theory to  study the
phonon-induced decoherence  on vacuum Rabi oscillations as a function of detuning between the cavity mode and exciton.
Kaer {\em et al.} \cite{KaerPRL:2010}  also explored off-resonant interactions between the QD and
a cavity
using a numerical solution to the system master equations within
a time-convolutionless approach.
Ota {\em et al.} \cite{OtaArxiv:2009} numerically solved
the Wilson-Rae and Imamogl{\u u}  system master equation \cite{WilsonRayPRB:2004},
and demonstrated the importance of asymmetric off-resonance coupling and non-Markovian relaxation,
finding good agreement with their experiments.
Hohenester \cite{HohenesterPRB:2010} introduced a useful model to derive an effective
phonon-mediated scattering rate (see also Ref.~\cite{JiaoJPC2008}), and has  found good agreement with experiments in the weak coupling regime~\cite{HohenesterPRB:2009}.

From an experimental viewpoint,
significant off-resonant coupling
between an exciton and a cavity
has
been seen in a number of semiconductor  QD-cavity systems, e.g.,
see Refs.~\cite{RuthOE:2007,HennessyNature:2007,KaniberPRB:2008,SufczynskiPRL:2009,TawaraOE:2009};
and  several theoretical and experimental works have tried to explain the basic
coupling mechanisms \cite{AuffevesPRA:2008,YamaguchiOE:2008,NaesbyPRA:2008,shughesOE:2009,WingerPRL:2009},
most of which stem from the exciton broadening mechanisms and the simple
physics of two coupled oscillators. Given the importance of
exciton decay processes on the QD-cavity coupling, one can therefore expect that
the interaction of phonons can play a qualitatively important role when certain
coupling conditions are met.
However,  while  the aforementioned theoretical phonon studies are interesting in their own right, unfortunately none of them
present simple analytical spectra that allow one to explore a wide
range of coupling phenomena both on- and off-resonance with a {\em strongly coupled} cavity.
 Thus many of the
experimental -- and theoretical -- groups continue to use the only available
analytical formulas with no phonon coupling directly included, e.g.,
see Ref.~\cite{Dalacu:PRB2010}.

Here  we  apply a photon Green function method  to derive useful analytical spectra, with phonons included to all orders through the exciton polarizability.
Although a similar semi-classical Green function approach was presented by Milde {\em et al.}~\cite{MildePRB:2007}, only on-resonance conditions were studied,  and the IBM was solved numerically. Tarel and Savona \cite{SavonaPRB:2010} have recently presented a semi-classical  Green function spectra with
phonons included  to second order (also previously presented in Ref.~\cite{MildePRB:2007}),
  where numerical solutions of the phonon baths were exploited.
We first study
leaky cavity emission with and without phonons and connect the results to the recent work  of Ota {\em et al.}~\cite{OtaArxiv:2009},
and find good qualitative agreement with their observations, namely
pronounced asymmetries for high energy or low energy cavity coupling
and an asymmetric vacuum Rabi doublet (at low temperatures, $T\approx4\,$K); we also demonstrate that phonon-induced
cavity suppression can occur, which is an effect that
stems from the real part of the phonon self-energy.
Secondly, we compare directly with recent experimental measurements by Dalacu {\em et al.} \cite{Dalacu:PRB2010},
who studied QD-cavity coupling in single site-controlled QDs and found that the data could not
be fit without adding in some unknown (detuning-dependent) cavity pump term;  in contrast, we  show that
this data can be well reproduced using our analytic model with no cavity pump term included at all.
Importantly, our  spectra can be applied to a wide range of systems, including
waveguides, and it rigorously applies to both weak and strong coupling regimes, and contains
 phonons to all orders (at the level of the IBM).
 We also show higher temperature results and systematically compare these with experimental data.
 In our experimental-theory analysis, we
include  two decay channels,  accounting for both
 the radiation-mode emission and the leaky cavity emission.

Our paper is organized as follows. In Sec.~\ref{sec:theory}, we introduce
the basic theory and analytical formulas for calculating the
emitted spectra from a strongly coupled QD-cavity system
in the presence of  electron-acoustic phonon coupling.
Example computed emission spectra with and without
phonon coupling are shown in Sec.~\ref{sec:results}.
Section \ref{sec:exps} compares our theory with recent
data on site controlled single QDs in photonic crystal cavities \cite{Dalacu:PRB2010}, and demonstrates
the significant influence of electron-phonon coupling.
In Sec.~\ref{sec:con}, we  conclude.

\section{Theory and analytical spectra} \label{sec:theory}

For this work,
we are interested in the linear spectrum and thus consider
a QD-cavity system that is  weakly pumped, {\em incoherently}, where the emission dynamics stem
from an excited electron-hole pair. In general, there have been several theoretical approaches
to this problem in the literature. Two of the most powerful methods
include Green function approaches \cite{shughesOE:2009}
and quantum master equation techniques  \cite{LaussyPRL:2008,arnePRL2009,YaoPRB:2010}.
A major advantage of the Green function approach is that one can obtain analytical spectra
for any inhomogeneous and lossy structures,
 including lossy metamaterial waveguides \cite{YaoMMPRB:2009} and a variety
of coupled cavity-waveguide systems \cite{YaoLPR:2010,YaoPRB:2009B}.
In contrast,
 a key advantage of the master equation approach is the ease with which it adds in additional
 dissipation effects such as pure dephasing, though at the
level of obtaining emission spectra, both Green function and master equation approaches can be equivalent.
A significant  disadvantage of the master equation approach is that it is typically limited to simple
leaky cavity systems with a Lorentzian decay, i.e., Markovian theory. Master equation solutions
can also consider more realistic initial conditions, such as those obtained
through steady-state pumping.

The general Green function approach to obtaining the
electric-field
operator has been described elsewhere \cite{wubs04,shughesOE:2009,YaoOE:2009}. Here
 we will briefly highlight the  theoretical background and concentrate
on presenting the general expressions for the field operator and the spectra.
At first we neglect non-radiative broadening on the exciton decay,
which allows us to obtain the exact analytical field operator without coupling to phonons.
Specifically,
we use a canonical Hamiltonian that quantizes the macroscopic
electromagnetic fields, and exploit the dipole approximation
for the QD-medium coupling:
\begin{eqnarray}
\hat H & & =  \hbar \omega_x \hat{\sigma}_x^+\hat{\sigma}_x^-+
\sum_\lambda\hbar \omega_\lambda
\hat{a}_\lambda^\dagger\hat{a}_\lambda \nonumber \\
& &
-i\hbar\!\!\sum_{\lambda}( \hat{\sigma}_x^- +
\hat{\sigma}_x^+)(g_{\lambda}
\hat{a}_\lambda-g_{\lambda}^*\hat{a}_\lambda^\dagger),
\end{eqnarray}
 where
$\hat a_\lambda$ represents the field mode operators and
\textbf{$\hat \sigma_x^{+/-}$} are the Pauli operators of the  QD
excitons, $\omega_{\lambda}$ is the
eigenfrequency corresponding to the transverse modes of the system
[$\mathbf{f}_\lambda(\mathbf{r})$], {\em excluding}  the dot;
$g_{\lambda}$ is the field-dot coupling coefficient,
defined through
$g_{\lambda}=\sqrt{\frac{\omega_\lambda}{2
\hbar\varepsilon_0}}\bm{\mu}_x \cdot
\mathbf{f}_\lambda(\mathbf{r}) $, with ${\bm \mu}_x={\bm n}_x
\mu_x $ the optical dipole moment of the exciton, aligned
along ${\bm n}_x$ (a unit vector).
 We consider only one {\em target exciton} in the spectral region of interest for the coupled QD
 and assume that coupling to the other polarized exciton is negligible.
The Heisenberg equations of motion for the operators can be
used to derive
the electric field operator~\cite{wubs04,shughesOE:2009}.
 Considering a weak excitation condition (i.e., we neglect higher-order
photon-correlation effects, which is valid in these systems for weak
powers~\cite{YaoPRB:2010}),
and assuming an excited QD in vacuum,
we derive the quantum field operator~\cite{shughesOE:2009}:
\begin{eqnarray}
\mathbf{\hat{E}}({\bf r},\omega) =
\frac{\G(\mathbf{r},\mathbf{r}_d;\omega)\cdot\mathbf{\hat{d}}_x(\omega)/\varepsilon_0}
{1- {\bf n}_x  \cdot \G({\bf r}_d,{\bf r}_d;\omega) \cdot {\bf n}_x \alpha_x (\omega)} ,
\end{eqnarray}
 where ${\bf r}_d$ is the QD position,  $\alpha_x(\omega)=\frac{\mu_x^2}{\hbar\varepsilon_0}\,\frac{2\omega_x}{\omega_x^2-\omega^2}$
 is the {\em bare} (no radiative or non-radiative coupling) exciton polarizability,
and  $\mathbf{\hat{d}}_x(\omega)  =
{-i\bm\mu_x}[ {\hat{\sigma}_x^-(t=0)}
/({\omega-\omega_x})+{\hat{\sigma}_x^+(t=0)}/({\omega+\omega_x})
]$ is a quantum dipole source that originates from the
excited QD.
Without phonon interactions, the above operator is exact within the stated
model approximations.
The propagator,
 $\mathbf{G}(\mathbf{r},\mathbf{r}';\omega)$,
is the transverse \cite{YaoPRB:2010} photon Green function of the medium, {\em without any QD},
and for cavity systems it can be easily written
down analytically in terms of the cavity mode
decay (cavity emission) and the radiation-mode decay (dot emission)
\cite{shughesOE:2009}.
Although electric field pumping  (e.g., cavity pumping) can be included
in a straightforward way, we are primarily interested
in the spectra from an excited (and single) QD, so we treat the
initial field as in {\em vacuum}.

With a knowledge of the medium-dependent Green functions,
and a suitable initial condition for exciting the
material system, one can conveniently obtain the analytic emitted spectrum, {\em in any general structure}, from~\cite{shughesOE:2009}
$S({\bf r},\omega)=\langle
[{\hat{\bf E}}({\bf r},\omega)]^\dagger{\hat{\bf E}}({\bf r},\omega)\rangle$.
The above expressions are exact, and no assumption has been made
about the form of the medium.

For a semiconductor cavity system such as a planar photonic crystal cavity,
one easily obtains the cavity emitted spectrum
and the radiation-mode emitted spectrum \cite{shughesOE:2009}, analytically,
\begin{eqnarray}
&&S_r({\bf r},\omega)= F_r({\bf r})\Gamma_{\rm rad} \times \nonumber \\
&\hbox{}&  \left | \frac{\omega_x+\omega}{\omega_x^2-\omega^2-i\omega\Gamma_x
-\frac{4g^2\omega_x\omega_c}{\omega_c^2-\omega^2-i\omega\Gamma_c}} \right |^2 , \ \ \ \  \\
&& S_c({\bf r},\omega) = F_c({\bf r})\Gamma_{c} \times \nonumber \\
&&  \left | \frac{ \frac{2g \omega_c (\omega_x+\omega)}{\omega_c^2-\omega^2-i\omega\Gamma_c}}{\omega_x^2-\omega^2-i\omega\Gamma_x
-\frac{4g^2\omega_x\omega_c}{\omega_c^2-\omega^2-i\omega\Gamma_c}} \right |^2 , \ \ \ \
\end{eqnarray}
with the total spectrum $S_t({\bf r},\omega)=S_r({\bf r},\omega)+S_c({\bf r},\omega)$.
We   emphasize that radiative
 coupling to the cavity system is fully included by coupling to both the
 continuum of radiation modes
and the leaky cavity mode---with a decay rate  given by $\Gamma_c$.
Additional broadening of the ZPL has also been included, phenomenologically, through
 $\Gamma_x \equiv \Gamma_{\rm rad}+\Gamma'$, with $\Gamma'$ due to
  pure dephasing processes; in general,
 for cw (continuous wave) pumping, both phonon effects
 and spectral diffusion will enhance $\Gamma'$,
 and $\Gamma'$ is also known to be temperature
 dependent \cite{BorriPRL:2001}.
It is important to note that
in a planar photonic crystal structure, both $S_r$ and $S_c$ photon decay channels
contribute to
vertical photon emission. The radiation-mode decay is due to coupling to
the continuum of radiation modes above the slab light line \cite{shughesOE:2009};
and
the $F_{c/r}({\bf r})$ represent the geometrical factors which will depend upon the
collection geometry of light emission.
For a micropillar cavity system \cite{pillars}, typically only the
cavity emission is required in an identical form to above, and so the prescriptions
apply to a wide range of semiconductor cavity systems.

To include phonon interactions in a simple but rigorous way,
we assume that the cavity and phonon correlation functions can be decoupled,
and add in the phonon polarizability via the known phonon self-energy, $\Sigma_{ph}(\omega)$, from the IBM
In essence, we are then considering
the optical polarizability of the QD in the presence of phonons, as the exact perturbation
to the medium.
Consequently, one
has a slight modification to the above spectra,
\begin{eqnarray}
&&S_r({\bf r},\omega)= F_r({\bf r})\Gamma_{\rm rad} \times \nonumber \\
&\hbox{}&  \left | \frac{\omega_x+\omega}{\omega_x^2-\omega^2-i\omega\Gamma_x -\omega\Sigma_{ph}(\omega)
-\frac{4g^2\omega_x\omega_c}{\omega_c^2-\omega^2-i\omega\Gamma_c}} \right |^2 , \ \ \ \  \\
&& S_c({\bf r},\omega) = F_c({\bf r})\Gamma_{c} \times \nonumber \\
&&  \left | \frac{ \frac{2g \omega_c (\omega_x+\omega)}{\omega_c^2-\omega^2-i\omega\Gamma_c}}{\omega_x^2-\omega^2-i\omega\Gamma_x -\omega\Sigma_{ph}(\omega)
-\frac{4g^2\omega_x\omega_c}{\omega_c^2-\omega^2-i\omega\Gamma_c}} \right |^2 , \ \ \ \
\label{Eq:final}
\end{eqnarray}
where the cavity emission is similar in form to the one presented by Tarel and Savona \cite{SavonaPRB:2010},  where  phonons were included to second-order
and a rotating-wave approximation was made.
As limits, we obtain the correct IBM spectral form
for exciton decay, and earlier derived spectra
for semiconductor cavities \cite{shughesOE:2009}. We also obtain
ZPL broadening associated with the leaky cavity system.


In the spirit of deriving a simple analytic solution,
with phonons included to all orders,
 the strategy is to use an analytic phonon self-energy
at the level of the IBM. To do this, we
exploit phonon spectral functions, similar to the ones used by
Wilson-Rae and Imamogl{\u u}  \cite{WilsonRayPRB:2004}, but we use
a more appropriate spectral function for
phonon interactions via a deformation potential \cite{CalarcoPRB:2003};
similar spectral function are commonly used when describing LA-phonon coupling.
To obtain the phonon self-energy, the IBM time-dependent phase must be added
into the Lorentzian decay model for the exciton; this is obtained from
\begin{equation}
 \psi(t) = \int_0^\infty d\omega J(\omega)/\omega^2 [\coth(\beta\hbar \omega/2)\cos(\omega t)
-i\sin(\omega t)],
\end{equation}
that describes the electron-LA-phonon interaction, at temperature $T=1/\beta k_b$;
 we consider electron-phonon interactions via a deformation potential coupling
which is known to account for the major phonon interactions in our considered QD.
Although one may have LO interactions as well, these give no dephasing.
We considering a spherical QD model,  with similar electron localization lengths
in the valence and conduction bands ($l_e=l_h \approx 5\,$nm).
Generalizing to include electrons and holes with different localization lengths
is straightforward, but the equations become more cumbersome. Thus,
spectral function can be conveniently defined as \cite{CalarcoPRB:2003}
\begin{equation}
J(\omega)=  a_p \, \omega^3 \exp(-\omega^2/2\omega_b^2),
\end{equation}
where we use $\omega_b=1\,$meV and $ a_{p}/(2\pi)^2=0.06\,{\rm ps^2}$ as representative
numbers for InAs type QDs \cite{JiaoJPC2008,params}.
The deformation coupling constant used here is somewhat smaller
than the value used in Ref.~\cite{OtaArxiv:2009} (where $ a_{p}/(2\pi)^2  \approx 0.1\,{\rm ps^2}$),
though in the literature there is no well accepted values
for InAs QDs; moreover, the
dimensionless Huang-Rhys, $S_{HR} = a_p/(2\pi)^2\, c_l^2/l_{e/h}^2 \approx 0.034$ (where $c_l=3800\,{\rm m/s}$ is the speed of sound),
has been shown to be significantly enhanced in QDs
(from its bulk value); for example,
for InAs/GaAs QDs,
$S_{HR} = 0.01-0.5$ (Ref.~\cite{BissiriPRB:2000})  and  $S_{HR} =0.5$ (Ref.~\cite{WolterPSS:1999}) have been reported,
and various mechanisms for such enhancements have been proposed,
including defects  and non-adiabatic effects.
The phonon interactions also result in a polaron shift, $\Delta = \int_0^\infty d\omega J(\omega)/\omega = S_{HR}\,\omega_b \sqrt{\pi/2}
\approx 42\,\mu$eV.
In what follows below,
we neglect the polaron shift   as it merely adds a fixed frequency, and we can always redefine
$\omega_x$, so our $\omega_x$ includes the
polaron shift.
We  also point out that there will likely be other
non-diagonal phonon coupling as well and so in principle
the phonon parameters above could  be varied and used to fit experiments.

\begin{figure}[hb]
\includegraphics[width=8.2cm]{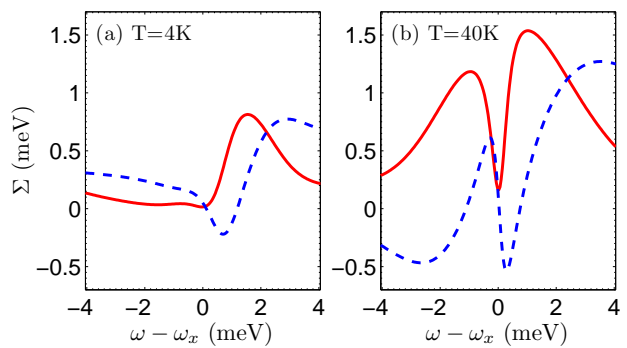}
\vspace{-0.4cm}
	\caption{\label{FigTh1}  (Color online)
Phonon self energies
for InAs QDs, for two different temperatures ($T=4,40\,$K),
where the red-dark curves represent the imaginary contribution
and the grey-light curves represent the real contribution.
The broadening parameters are
$\Gamma_{\rm rad}=2\,\mu$eV and $\Gamma_x'=75\,\mu$eV.
As discussed in the text, $\omega_x$ is considered to
already include a polaron shift.
}
\end{figure}

With the above analytical form for the phonon bath,
the time-dependent polarizability takes the form $\alpha_x(t) = \alpha_x(0)\exp[-i(w_x+\Delta)t+\psi(t) ]$,
and the frequency dependent polarizability is obtained from a simple Fourier transform, yielding
\begin{equation}
\alpha_{x}(\omega) = \frac{d^22\omega_x/\hbar\varepsilon_0}{\omega_x^2-\omega^2-i\omega\Gamma_x -\omega\Sigma_{ph}(\omega)}.
\end{equation}
To help better explain the phonon coupling effects shown later, in Fig.~\ref{FigTh1} we
show the phonon self-energies for the two different temperatures
of $T=4\,$K (a) and $T=40\,$K (b). We recognize a local minimum in ${\rm Im}[\Sigma_{ph}]$ (red-solid)
near $\omega_x$, a maximum in ${\rm Im}[\Sigma_{ph}]$ near 1\,meV, and significant
${\rm Re}[\Sigma_{ph}]$ (blue-dashed) over a broad spectral range. These results imply that both the real and imaginary
contributions will have a significant impact on phonon coupling effects
in the regime of cavity-QED.

%
%
%


As mentioned above, our general methodology is similar in spirit to a semi-classical approach. Specifically,
we have assumed a well defined spectral lineshape for the QD susceptibility, and then coupled this QD frequency response -- with phonons included self-consistently --
to the medium-dependent Green functions to obtain the analytical spectra.
Comparing with the approach of Wilson-Rae and Imamogl{\u u}, they have
in their system Hamiltonian a phonon-modified cavity coupling rate \cite{WilsonRayPRB:2004} $g \rightarrow g \braket{B}$, with $\braket{B} = \exp(-0.5 \int_0^\infty d\omega J(\omega)
\coth(\beta\hbar\omega/2)/\omega^2)$, which, using the parameters above
for $T=4-40\,$K, is around $\braket{B}=0.91-0.55$.
We do not have this term explicitly, however
our self-energy term naturally includes this coupling, and to all orders. To make this clearer, we can
rewrite the solution as, e.g., for the cavity-mode emission,
\begin{eqnarray}
%
%
&& S_c({\bf r},\omega) = F_c({\bf r})\Gamma_{c} \times \nonumber \\
&& \!\!\!\!\!\! \left | \frac{ \frac{2g \omega_c (\omega_x+\omega)}{\omega_x^2-\omega^2-i\omega\Gamma_x}
\left ( 1+\frac{\omega\Sigma_{ph}}{\omega_x^2-\omega^2-i\omega\Gamma_x} + \cdots \right )  }
{\omega_c^2-\omega^2-i\omega\Gamma_c
-\frac{4g^2\omega_x\omega_c}{\omega_x^2-\omega^2-i\omega\Gamma_x}
\left (1+\frac{\omega\Sigma_{ph}}{\omega_x^2-\omega^2-i\omega\Gamma_x} + \cdots \right )
} \right |^2 , \nonumber \\
&&
\end{eqnarray}
where the real part of the phonon self-energy causes a reduction in the
$g$ coupling. Such an effect was previously pointed out before for
on-resonance coupling between a cavity and an exciton \cite{MildePRB:2007}.

%

\section{Theoretical Simulations and Predictions}
\label{sec:results}

We first clarify that with no phonon coupling included, then
our Green function technique yields {\em identical} normalized spectra to a master equation
approach in the low power limit. Specifically,
we use the equations in Ref.~\cite{YaoPRB:2010}, with a low-power incoherent exciton pump, and
we compute the
total spectra from both cavity emission and QD emission
in the presence of pure dephasing; with the Green function
approach, we define
$\Gamma_{ZPL}=\Gamma_{\rm rad}+\Gamma'$.
These spectral forms are found to be identical, as
previously discussed \cite{shughesOE:2009}, so the general
phenomenon of cavity feeding is not unique
to pure dephasing processes. The only difference
is the overall magnitude, which is due to the different
initial conditions, though these can be made equivalent as well.
Any exciton broadening (radiative or non-radiative) will feed the cavity mode in the cavity-mode emission,
and scale with $g^2$.
 However, the distinction of {\em pure dephasing} plays a much more important role for computing
higher-order quantum correlations \cite{Emy}.

\begin{figure}[t!]
\includegraphics[width=8.6cm]{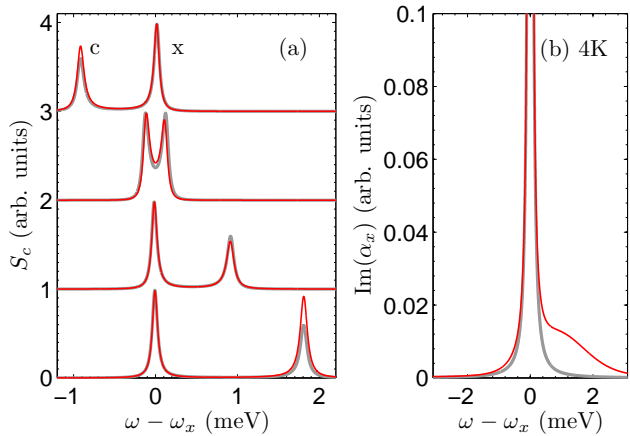}
\vspace{-0.5cm}
	\caption{\label{FigTh2}  (Color online) (a)
Cavity-emitted spectra for various QD-cavity detunings, obtained for a
temperature of
$T=4\,$K.
The  red-dark  and grey-light curves display the
Green function
solution with and without LA-phonon interactions.
The corresponding imaginary part of the polarizability with (red-dark curve) and without (grey-light curve) phonon coupling is shown
in (b).
The lower frequency
phonon bath is suppressed  in part due to the low temperatures and in  part due to
the relatively large ZPL.
 The polaron shift, $\Delta \sim 30\,\mu$eV,
 is not included
as this just adds in a fixed resonance shift for all detunings and temperatures.
The parameters are $\Gamma_{\rm rad}=2\,\mu$eV, $\omega_x = 830\,$meV, $\Gamma_x'=75\,\mu$eV,
$\Gamma_c=\,100\,\mu$eV, and $g=0.13\,$meV (see Ref.~\cite{cavParams}).
}
\end{figure}
%


In Fig.~\ref{FigTh2}(b), we show the imaginary part of the QD
polarizability (which is proportional to the absorption) with and without
coupling to phonons via the IBM; this calculation was obtained at a temperature
of 4\,K, and we clearly obtain the familiar spectral form
of the IBM  spectral lineshape \cite{KrummheuerPRB:2001,VagovPRB:2002,KnorrPRL:2003}.
The phonon-induced lineshape is subsequently used to obtain the modified spectra
with phonon coupling.
In Fig.~\ref{FigTh2}(a) we plot the Green function cavity spectra solution with (red-dark curve) and without
(grey-light curve) phonons, where the QD and cavity parameters
are given in the figure caption.
At zero detuning,  we recover
an asymmetric Rabi splitting in agreement
with Refs.\cite{MildePRB:2007,OtaArxiv:2009}; in addition, for
small detunings, a positive detuning gives
a cavity {\em suppression}, while a negative detuning gives
a cavity enhancement.
These are due to the significant frequency shifts from the
real part of the phonon self-energy,  whose effects would be absent in
 an effective rate approximation.
This suppression, followed by an enhancement, would
exacerbate the effects of cavity coupling for positive
detunings, and depend upon the spectral form of the
phonon bath.
 The off-resonant
cavity peak primarily stems from $S_c$ and scales with $g^2$.

It is also worth stressing that the asymmetric vacuum Rabi doublet is an effect
that cannot be predicted with a standard master equation approach, nor by
 using an effective cavity feeding rate~\cite{HohenesterPRB:2009}.
As pointed out some time ago by Carmichael and Walls~\cite{carmichael},
the usual decomposition of the system Hamiltonian to include only the noninteracting QD and cavity parts
does not satisfy the detailed balance condition~\cite{carmichael}. This is due to the fact that the cavity and the QD systems are {\em internally coupled},
and modified master equation must be derived to account for this internal coupling.
  In fact, the polaron transformation used
  by Wilson-Rae and Imamogl{\u u} \cite{WilsonRayPRB:2004} adopts such an approach, and so their
   transformed system Hamiltonian leads to the correct form of the density operator
   -- since it preserves detailed balance conditions~\cite{carmichael}.
   Physically the asymmetric vacuum Rabi splitting~\cite{MildePRB:2007},
  which has recently been measured \cite{OtaArxiv:2009}, comes form the fact that the two coupled
   Rabi peaks sample different parts of the asymmetric phonon bath.



\begin{figure}[t]
\includegraphics[width=8.6cm]{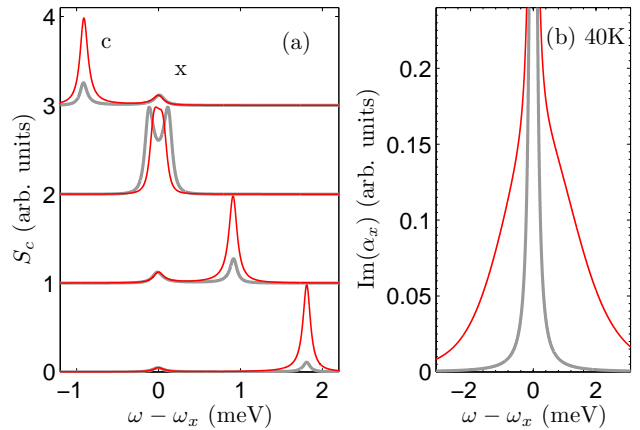}
	\caption{\label{FigTh3}  (Color online)
As in Fig.~\ref{FigTh2}, but with  $T=40\,$K
and $\Gamma'=150\,\mu$eV.
}
\end{figure}

Since we include phonons to all orders, it is also possible to study elevated
temperatures, which is highly desired from a device viewpoint. In Fig.~\ref{FigTh3}  we repeat the same calculation as before, but at
$T=40\,$K (chosen primarily to connect to the experiments below), and the ZPL has been broadened (doubled) to be consistent with experiment \cite{BorriPRL:2001}.
Again, we see the same phonon coupling  trends as before, but now with a significantly
larger phonon feeding to the cavity mode.
Also, the effect of the cavity coupling parameters is significantly reduced by about 50\%,
resulting in a vanishing
Rabi splitting on resonance when phonons are included.
All of these predicted trends are consistent with the recent
experiments of Ota {\em et al.}\cite{OtaArxiv:2009}.


\section{Comparison with recent experimental data on a single QD- photonic crystal
cavity system}
\label{sec:exps}

 We now apply our theory to help explain the recent model {\em discrepancies}
that were employed for simulating experimental data.
Earlier this year, Dalacu {\em et al.}~\cite{Dalacu:PRB2010} reported measurements on a single QD-cavity system at low power
and, using a master equation theory of Refs.~\cite{YaoPRB:2010,TawaraOE:2010}, and clearly demonstrated that an extra cavity pump term
had to be included {\em by hand} to explain their data. The cavity pump contribution was found
to be
cavity-QD detuning-dependent. The origin of this cavity pump was unknown and cited to be somewhat mysterious for their excitation
conditions and sample,
but similar couplings have been shown already to be due to
electron-phonon coupling \cite{JiaoJPC2008,HohenesterPRB:2009,OtaArxiv:2009}.
The site-controlled QD allows one to suppress other extraneous cavity feeding mechanisms \cite{KaniberPRB:2008,WingerPRL:2009} , and
for detunings greater than 5\,meV,
they observe no cavity-mode emission at low temperatures; this is
 expected for a single exciton-cavity coupling \cite{shughesOE:2009}, since the
cavity-mode emission diminishes as a function of detuning.

\begin{figure}[b]
\includegraphics[width=7cm]{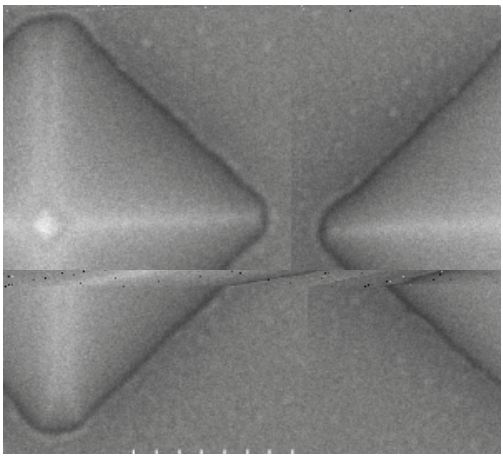}
\caption{A single InAs quantum dot nucleated at the apex of a InP pyramid. The pyramid is grown using selective-area epitaxy on a electron-beam patterned SiO$_2$-coated InP substrate. The scale bar is 210\,nm.}\label{pyramid}
\end{figure}

The single QD-cavity system is realized by nucleating one InAs QD at the apex
 of a InP pyramid (see Fig.~\ref{pyramid}) grown using selective-area epitaxy\cite{Poole_NT10}.
 Dot formation on these InP pyramidal nanotemplates proceeds via
 the Stranski-Krastanow growth mode\cite{SK} similar to growth on planar substrates although some subtleties of the strain distribution will differ due to the proximity of the $\{110\}$ planes that make up the sidewalls of the InP pyramid.  Once planarized, however, any signature of the InP pyramid vanishes, leaving a coherent InAs dot in a uniform InP matrix.
One distinction between planar and site-controlled dots pertains to the wetting layer. Although the presence of a wetting layer is assumed, its lateral extent is limited to the apex of the pyramid and is thus only slightly larger than the QD. The site-controlled dots are therefore expected to have a modified shape compared to planar dots, and the absence of an infinite 2D wetting layer. Although these differences to do not manifest in the electronic structure of the dots~\cite{Kim_PSSC06}, the absence of a wetting layer and associated continuum of states may have implications related to non-resonant dot-cavity coupling~\cite{WingerPRL:2009}. Clearly, reducing unknown and unwanted
excitation mechanisms should be avoided from a practical perspective,
which makes dots without a wetting layer advantageous, e.g.,
for creating cavity-assisted single photons on demand
\cite{Pathak_2010}.

\begin{figure}[t!]
\includegraphics[width=8.5cm]{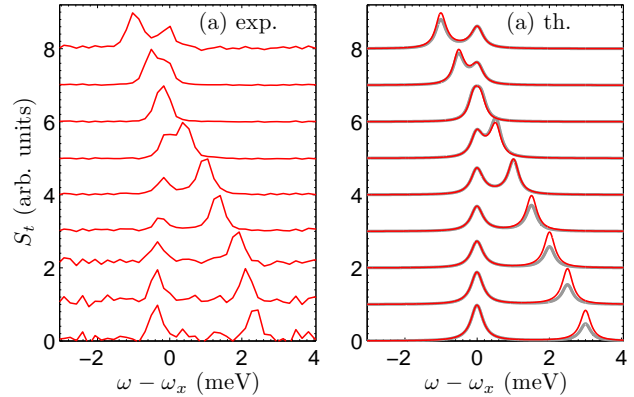}
\vspace{-0.2cm}
	\caption{\label{FigExpTh1}  (Color online) (a)
Recently published experimental data (Dalacu {\em et al.}~\cite{Dalacu:PRB2010}), taken
at $T=4\,$K,
normalized to show the effect of detuning on the exciton and cavity mode.
(b) Theoretical simulations with (red-dark) and without (grey-light) phonon interactions.
The parameters are similar to those used in Fig.~(1), except
that $\Gamma'=150\,\mu$eV and we have convolved
with a Lorentzian function with FWHM $\Gamma_{\rm spec}=250\,\mu$eV to account for the spectral resolution
in the experiment \cite{Dalacu:PRB2010}. The detunings are not exactly fit to experiment, but rather are chosen
to cover a similar range to the experiments with equal frequency spacing. Both sets of calculations
are normalized to their peak value for clarity. We also
include the radiation-mode emission, with $F_c=2 F_r$.
}
\end{figure}

In Fig.~\ref{FigExpTh1}(a) we show the data of Dalacu {\em et al.}~\cite{Dalacu:PRB2010},
 and in Fig.~\ref{FigExpTh1}(b) we show our analytical model with (red-dark curve) and without (grey-light curve) phonon coupling. Evidently, the analytic phonon model well reproduces
the experiment data without the need to artificially add in a cavity pump term, and the phonon coupling
plays a  role on the cavity feeding mechanism. This general conclusion
is consistent with earlier works \cite{HohenesterPRB:2009,OtaArxiv:2009,KaerPRL:2010,SavonaPRB:2010}, though some of the qualitative trends are quite different (e.g., a cavity suppression followed by
an enhancement); moreover,
since
our final spectral forms are analytic,
 one can sweep a wide range of parameters
in a straightforward and simple way.
Moreover, our formulas can be applied to fit spectra over
a wide range of temperatures (up to room temperature), where one must go beyond a second-order
approximation for the phonon baths.

\begin{figure}[t!]
\includegraphics[width=8.5cm]{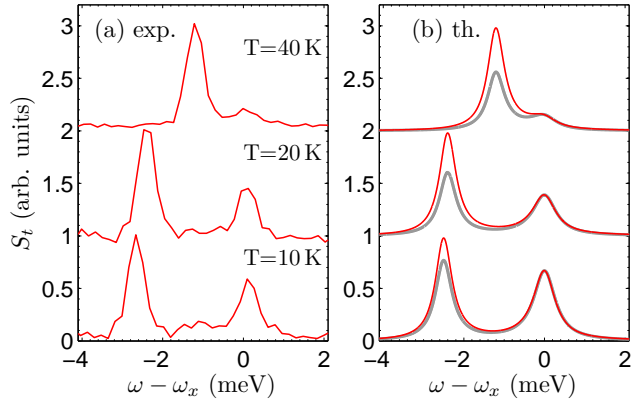}
	\caption{\label{FigExpTh2}  (Color online) (a)
Measured spectra at various temperatures, taken
at $T=10\,$K,  $T=20\,$K, and $T=40\,$K.
(b) Theoretical simulations with (red-solid curve) and without (grey-light curve) phonon interactions.
The parameters are similar to those used in Fig.~5, except
that $\Gamma'=225-300\,\mu$eV (increasing linearly with temperature) and we have convolved
with a Lorentzian function with FWHM $\Gamma_{\rm spec}=400\,\mu$eV \cite{DanComment}.
}
\end{figure}

Finally,  we fit the emission spectra for elevated temperatures, using data that has not previously
been shown. The sample and experiment are
similar to those described by Dalacu {\em et al.}~\cite{Dalacu:PRB2010}.
In Fig.~\ref{FigExpTh2} we show the experimental data (a) and simulations (b)
for a temperature-controller  cavity detuning,
for temperatures of 10\,K, 20\,K, and 40\,K.
Once more, we  see a good trend with the experiments, and the effect
of the phonons is to enhance the oscillator strength of the cavity mode relative
to the exciton mode, which becomes stronger as a function of temperature.\\

\section{Conclusions}
\label{sec:con}

We have presented a Green function theory to describe photon emission in arbitrary QD-cavity systems,
without recourse to either the weak coupling regime or a perturbative approximation for the phonons.
We have exploited this approach to model the strong coupling regime between a single exciton and
a photonic crystal cavity, and, using
a very simple initial condition of an inverted electron-hole pair, obtained the emission spectra for various dot-cavity detunings. Phonon related effects
such as cavity-mode suppression and enhanced cavity feeding are demonstrated, showing the need
to include both real and imaginary phonon self-energy terms with internal coupling.
 The model was then applied to help explain
 experimental data as a function of cavity-exciton detuning and as a function of temperature;
good agreement has been found without artificially adding in a cavity-mode
pump---as was done previously \cite{Dalacu:PRB2010}.
The phonon-induced cavity coupling thus mimics an incoherent cavity pump term,
which naturally would be detuning dependent because of the asymmetric phonon bath.
Our results and general predictions are also consistent with the recent data
and polaron master equation \cite{WilsonRayPRB:2004} simulations of Ota {\em et al.} \cite{OtaArxiv:2009}.

\section*{Acknowledgments}
This work was supported by the
National Sciences and Engineering Research Council of
Canada and the Canadian Foundation for Innovation.
We gratefully acknowledge U. Hohenester,
Y. Ota, and C. Roy for useful discussions.

\end{document}